\documentclass[epsfig]{article}
\usepackage{graphicx}
\begin{document}
\title{Time-of-arrival probabilities and quantum measurements: III Decay of unstable states}
\author{Charis Anastopoulos\footnote{anastop@physics.upatras.gr}\\
{\small Department of Physics, University of Patras, 26500 Patras,
Greece}} \maketitle

\begin{abstract}
We study the decay of unstable states  by formulating quantum
tunneling as a time-of-arrival problem: we determine the detection
probability for particles at a detector located a distance $L$ from
the tunneling region. For this purpose, we use a
Positive-Operator-Valued-Measure (POVM) for the time-of-arrival
determined in \cite{AnSav06}. This only depends on the initial
state, the Hamiltonian and the location of the detector. The POVM
above provides a well-defined probability density and an unambiguous
interpretation of all quantities involved.  We demonstrate that the
exponential decay only arises if  three specific mathematical
conditions are met. Their physical content is the following: (i) the
decay time is much larger than any microscopic timescale, so that
the fine details of the initial state can be ignored, (ii) there is
no quantum coherence between the different `attempts' of the
particle to traverse the barrier, and (iii) the transmission
probability varies little within the momentum spread of the initial
state. We also determine the long time limits of the decay
probability and we identify  regimes, in which the decays have no
exponential phase.

\end{abstract}

\renewcommand {\thesection}{\arabic{section}}
\renewcommand {\theequation}{\thesection. \arabic{equation}}
\let \ssection = \section \renewcommand{\section}{\setcounter{equation}{0} \ssection}

\section{Introduction}

This is a third in a series of papers \cite{AnSav06, AnSav07a},
which studies the properties and applications of a
Positive-Operator-Valued-Measure for the time-of-arrival. In
\cite{AnSav06} this POVM was constructed and it was applied to the
free particle case, where it coincided with the one of Kijowski
\cite{Kij74}. In \cite{AnSav07a}, this POVM was adapted to the
tunneling case: it led the determination of tunneling time and
through a generalization to sequential measurements it provided a
probability distribution for the times of arrival. Here, we apply
the formalism for the determination of the decay probability of
unstable quantum states through tunneling.

Quantum tunneling refers to the escape of a particle from a region
 through a potential barrier, whose peak is larger
than the particle's energy. The question we answer here is
 what is the law that determines the rate of the
particle's escape through the barrier?

  The issue of the escape probability for particles through a barrier as a
function of time has been  important  ever since the first days of
quantum mechanics. It is an observed fact that in the vast majority
of physical systems the escape rate is approximately constant: the
decay is exponential. However, the constancy of the escape rate does
not hold at very early \cite{nonexp1} and at very late times
\cite{nonexp2}. In the first regime, there is a behavior
corresponding to quantum Zeno effect and in the latter the decay is
governed by an inverse-power law (perhaps with oscillations).
Moreover, it is not necessary that  all systems decay exponentially.
The method we develop in this paper allows a full characterization
of the decay behavior for an unstable state that decays through
tunneling.

\subsection{Our approach}

The basic feature of our approach to both problems is its
operational character. We identify  the escape probability by
constructing probabilities for the outcome of specific measurements.
We assume that the quantum system is prepared in an initial state
$\psi_0$, which is localized in a region on one side of a potential
barrier that extends in a {\em microscopic} region. At the other
side of the barrier and a {\em macroscopic} distance $L$ away from
it, we place a particle detector, which records the arrival of
particles. Using an external clock to keep track of the time $t$ for
the recorder's clicks, we construct a probability distribution
$p(t)$ for the time of arrival. The fact that the detector is a
classical macroscopic object and that it lies at a macroscopic
distance away from the barrier allows one to state (using classical
language) that the detected particles must have passed through the
barrier (quantum effects like a particle crossing the barrier and
then backtracking are negligible). Hence, at the observational
level, the probability $p(t)$ contains all information about the
temporal behavior for the ensemble of particles.

For the purposes of this paper, we assume that each particle in the
ensemble is located within a microscopic region, which is bounded by
a potential barrier (e.g. nucleus). For the particle to escape this
region and to be detected at a macroscopic distance away,  it must
tunnel through the barrier. In effect, the initial state is unstable
and it decays through tunneling. It is evident that the detection
probability $p(t)$ incorporates the decay probability for the
unstable state: modulo a transient period (before the first
particles arrive at the detector), its physical content is the same.

 With the considerations above, the problem of determining the escape probability for a state that decays through tunneling
  is equivalent to the determination of probability for the time-of-arrival
  for an ensemble of particles described by the wave function
$\psi_0$ at $t= 0$ and evolving under a Hamiltonian with a potential
term. To solve this problem, we use the result of \cite{AnSav06},
namely the construction of a POVM for the time-of-arrival for
particles for a generic Hamiltonian $\hat{H}$.
 This POVM provides a unique
determination of the probability distribution $p(t)$ for the
time-of-arrival. It is important to emphasize that by construction
$p(t)$ is {\em linear} with respect to the initial density matrix,
positive-definite,  normalized (when the alternative of
non-detection is also taken into account) and a genuine density with
respect to time. We refer the reader to \cite{AnSav06} for the
physical assumptions relevant to the definition of this POVM and to
\cite{AnSav07a} for its detailed construction for the case relevant
to tunneling.

Our approach involves the formulation of the decay probability as a
time-of-arrival problem. There are other approaches that treat
quantum tunneling as a time-of-arrival problem in the literature. In
Ref. \cite{AHN00}, a distribution for tunneling time is obtained by
considering a detector model and defining a probability distribution
for time as $P_X(t) = \frac{ |\psi(X,t)|^2}{\int_0^{\infty} ds
|\psi(X, s)|^2}$, where $X$ is the position variable.  The method
yields positive definite probabilities. Unlike the present
treatment, these probabilities are not linear with respect to the
initial state (density matrix). Our approach is closer to the one
developed in
 Ref. \cite{HSMN}, in which a method is developed that models the measurement
with an imaginary potential near the detection point--see also
\cite{HSMN2}. This leads to a class of POVMs for the time-of-arrival
that provide a generalization of Kijowski's and are similar in form
to the one we employ in this paper.

\subsection{Comparison to other approaches }

The study of the decay probability of unstable states goes back to
the first days of quantum mechanics, most notably in the work of
Weisskopf and Wigner \cite{WW}. It is known that for  the
(overwhelming) majority of physical systems the decay laws of
unstable states are exponential. It is also known that the
exponential law cannot be valid at very short and at very long
timescales. This implies that even though the exponential decay law
is very common, it is not universal. This fact brings about many
questions: why  is the exponential decay observed valid in such a
variety of systems? which are the physical conditions necessary for
its appearance? are there systems, whose decay has no exponential
phase?  We shall see that the probability distribution $p(t)$ allows
us to provide an answer to these questions (at least for decays that
can be formulated as a tunneling problem).

Most studies of the validity of the exponential decay law (see in
particular \cite{Ghi, MCS78, Per80, GV}) proceed through the
determination of the properties of the survival amplitude $\langle
\phi_0|e^{-i\hat{H}t} |\phi_0 \rangle$, where $\phi_0$ is the
initial (unstable) state and $H$ the system's Hamiltonian. The
modulus square of the amplitude is the probability that the system
lies at the state $|\phi_0 \rangle$ at time $t$. The exponential
decay then refers to the behavior of this probability.

The survival probability is a function of time: however, it is not a
probability {\em density} with respect to time. One can immediately
see this on dimensional grounds: $|\langle \phi_0|e^{-i\hat{H}t}|
\phi_0 \rangle|^2$ is a pure number, while a genuine probability
density has dimensions of inverse time. However, the quantity $w(t)
:= 1 - |\langle \phi_0|e^{-i\hat{H}t}| \phi_0 \rangle|^2$ is the
total probability that the state $\phi_0$ has decayed at time $t$.
If $|\langle \phi_0|e^{-i\hat{H}t}| \phi_0 \rangle|^2 \rightarrow 0$
as $t \rightarrow \infty$, the first derivative of $w(t)$ can be
interpreted as a normalized probability density $p(t)$ for the decay
at time $t$. This argument would work in classical probability
theory. However, in quantum theory
\begin{eqnarray}
\dot{w}(t) = - i Tr \left( \rho_0 [ \hat{H}, \hat{Q}(t)] \right),
\end{eqnarray}
where $\hat{\rho} = |\phi_0 \rangle \langle \phi_0|$ and $\hat{Q}(t)
= e^{i\hat{H}t} |\phi_0 \rangle \langle \phi_0| e^{-i\hat{H}t}$. The
quantity $-i[\hat{H}, \hat{Q}(t)]$ is not a positive operator, hence
there is no guarantee that $\dot{w}(t)$ will be positive at all
times for a generic initial state $t$. Its interpretation as
probability density is therefore problematic.

Clearly, for exponential decay $\dot{w}(t) > 0 $. However, the
exponential decay does not hold at all moments of time, and the
survival probability is not a monotonously decreasing function of
time. Hence, $\dot{w}(t)$ does not define a probability distribution
for all $t \in [0, \infty)$. For the times that the exponential
decay holds, it is reasonable (if ultimately unjustified) to think
of $\dot{w}(t)$ as a probability density, but outside the
exponential regime this interpretation cannot be trusted. In fact,
if one views $\dot{w}(t)$ as a functional of the initial state
$\rho$, it is not a linear functional; hence, convex combinations of
initial density matrices do not lead to convex combinations of
`densities' $\dot{w}(t)$. This is a serious problem because
according to the usual interpretations of the quantum state, a
convex combination of initial states can  be achieved by joining
different statistical ensembles.

Another problem with the probabilities defined through the survival
amplitude is that they lack clear interpretation in terms of quantum
measurement theory. The `decay probability density' $\dot{w}(t)$ is
obtained formally as the expectation value of the operator
$-i[\hat{H}, \hat{Q}(t)]$: there is no clear procedure through which
this can be directly measured. Moreover,  the initial state (hence
also $\hat{Q}$) is often unknown. In general, there is no immediate
relation between $\dot{w}(t)$ and concrete measurement procedures
performed on a quantum system.

\subsection{Our results}

In the operational approach we follow here, the problems
characterizing the survival probability method do not arise. The
probability density $p(t)$ is genuine and the corresponding POVM
refers to a concrete procedure for the measurement of the decay
probability: we measure the time-of-arrival of the emitted
particles, and we construct a probability density that refers to an
ensemble of decaying quantum systems (e.g. nuclei). This method
allows for no interpretational ambiguities: the probabilities we
employ in this paper are soundly tied to the statistics of detection
outcomes.

We find that  exponential decay is generic, and we identify a
primary physical reason for its validity. In effect, the exponential
decay (at least in tunneling) arises when a specific condition holds
(this was first assumed by Gamow \cite{Gam} and by Gurney with
Condon \cite{LoGu} in their classic studies of alpha decay). In
semiclassical language, this condition is that the different
attempts of the bound particle to cross the barrier are {\em
statistically independent}, or in other words, that there is neither
interference nor memory effects in the probability distribution. The
exponential decay is then a sign of a `quasi-classical' (and
Markovian) behavior of the bound particle.

A second condition is that the decay time should be substantially
larger than the characteristic microscopic time scales associated to
tunneling: if this condition does not hold, the decay probability
exhibits a fine structure that is very sensitive to minor features
of the initial state. In effect, the validity of the exponential
decay law requires a separation of timescales, a condition similar
to the one for the Markovian behavior in open systems. In fact, in
absence of such a separation of time scales, it is questionable even
if the word "decay" is suitable for the description of the
phenomenon. A third condition is that there should be no coherence
between decays characterized by substantially different
characteristic timescales. We also show that the exponential phase
of the decay has a finite duration, and we find the asymptotic
behavior of the detection probability as $t \rightarrow \infty$
(inverse power law).

Our method provides a full characterization of the possible decay
laws for tunneling systems, at least for the class of potentials we
study here. (The choice of studied systems is guided by our desire
to obtain analytic expressions for all relevant quantities. However,
 the POVM we employ is defined for a generic Hamiltonian: hence, the
domain of applicability of the method is larger.) We find that there
are specific regimes in many systems, in which the decays have no
exponential phase. These regimes are identified by conditions on the
transmission and reflection amplitude of the barrier at the relevant
timescales.

The structure of this paper is the following. In Sec.2, we briefly
review the basic object in the formalism, namely the POVM
constructed in \cite{AnSav06, AnSav07a}. In Sec.3, we present an
informal argument about the assumptions involved in the derivation
of the exponential decay law and the physical conditions that these
necessitate. In Sec. 4, we construct explicitly the detection
probability for a state decaying through tunneling: we use a simple
model of a particle in the half-line, bounded close to $r = 0$ by a
potential barrier $V(r)$. In Sec. 5, we identify the regime of
exponential decay and we analyze the relevant conditions. In Sec.6,
we study the deviations from exponential decay at long times and
regimes for which the decays have no exponential phase. In Sec. 7,
we compare our results to the ones obtained from the survival
amplitude for the same systems and in Sec. 8, we conclude.

\section{Summary of the formalism}

The POVM defined in \cite{AnSav06, AnSav07a} refers to the following
situation. (We consider the case of a particle in one-dimension for
concreteness.) An ensemble of particles is prepared in a state
described by a wave function $\psi_0(x)$, which has support on
values $x < L$. The Hamiltonian is $\hat{H} = \frac{\hat{p}^2}{2M} +
V(\hat{x})$, where $M$ is the particle's mass. At $x = L$ a detector
is located, which register particles and records the time $t$ of
this recording. The distribution of these times-of-arrival is given
by a probability distribution
\begin{eqnarray}
p(t) = Tr\left( \hat{\rho}_0 \hat{\Pi}(t) \right), \label{probabdef}
\end{eqnarray}
where $\hat{\rho}_0 = | \psi_0 \rangle \langle \psi_0|$. The POVM is
defined through the operators $\hat{\Pi}(t)$, together with the
operator $\hat{\Pi}(N) = 1 - \int_0^{\infty} dt \hat{\Pi}(t)$, which
corresponds to the event of no detection. The sample space for this
POVM is therefore $[0, \infty) \times \{N\}$.

The POVM $\hat{\Pi}$ involves in its definition a smearing function
$f^{\tau}(t)$, which determines the response of the detector; $\tau$
is the characteristic response time. For the physically relevant
class of initial states that have support to values of energy $E$
such that $E \tau >> 1$, the POVM becomes $\tau$-independent. We
proceed to describe its structure.

For the calculation of the probability density (\ref{probabdef}), it
is necessary to find the spectrum of the Hamiltonian operator
$\hat{H}$ with Dirichlet boundary conditions at $x = L$. We assume
that the potential is short range, so that it vanishes around in the
neighborhood of $x = L$. (In fact, it is only different from zero in
a microscopic scale, while $L$ is a macroscopic distance. We
distinguish two cases: (i) if $x$ takes value in the half-line, the
spectrum of $\hat{H}_D$ is expected to be discrete (this is the case
relevant to this paper); (ii) if $x$ takes values in the full real
axis, at least the positive energy spectrum will be continuous.
Either way, for $x
>> a$, $V(x) = 0$ and the solution of the Schr\"odinger equation
$\hat{H}_D \psi_E(x) = E \psi_E(x)$ with Dirichlet boundary
conditions is proportional to $\sin k(L -x)$, where $k =
(2ME)^{1/2}$. We choose to label the eigenstates of $\hat{H}_D$ by
$k$, namely we write $|k\rangle_D$ as a solution to the equation
\begin{eqnarray}
\hat{H} |k \rangle_D = \frac{k^2}{2M} | k \rangle_D,
\end{eqnarray}
with Dirichlet boundary conditions.

Normalizing $|k \rangle_D$ so that
\begin{eqnarray}
{}_D\langle k| k' \rangle_D = \delta (k, k'),
\end{eqnarray}
(and similarly in the discrete-spectrum case) we write
\begin{eqnarray}
 \langle x|k\rangle_D = D_k \sin k (L-x),
\end{eqnarray}
where the form of the normalization factor $D_k$ is specified the
Hamiltonian's (generalized) eigenstates.

The probability distribution (\ref{probabdef}) is expressed as [see
Sec.2 in Ref. \cite{AnSav07a}]
\begin{eqnarray}
p(t) = \frac{1}{ 2\sqrt{2 }M} \sum_{kk'} D_k D_{k}^* c_k c^*_{k'}
\frac{kk'}{\sqrt{k^2 +k'^2}} e^{-i \frac{k^2 - k'^2}{2M} t},
\label{probab}
\end{eqnarray}

where $c_k = {}_D\langle k|\psi_0 \rangle$ and $\sum_k$ denotes the
integration with respect to the spectral measure of $\hat{H}_D$. The
probability for the time-of-arrival  is expressed solely in terms of
the system's Hamiltonian, the initial state and the value of $L$.

 Eq. (\ref{probab})  is
simplified if the spread $\Delta k$ of the initial state
$|\psi_0\rangle$ ( $\hat{k} = \sqrt{2M\hat{H}_D}$) is much smaller
than the corresponding mean value $\bar{k}$: in this case, $k^2 +
k'^2 \simeq 2 kk'$, hence
\begin{eqnarray}
p(t) =  \left| \sum_k D_k c_k \sqrt{\frac{k}{4M}}
e^{-ik^2t/2M}\right|^2.
\end{eqnarray}

\section{The origin of exponential decay}

For the study of the decay probability undertaken in this paper, we
will assume  a particle in the half line, described by a wave
function $\psi(r), r \in (0 ,\infty)$. The particle is initially in
region I ($0 < r < a $): the potential there can be in general
attractive, even though for simplicity we consider the case that
$V(r) = 0$. In region II ($a \leq r \leq b $), the potential $V(r)$
is repulsive and in region III ($ r
> b$), it vanishes.

Systems such as the above are described by an exponential law for
the decay probability. However, it is well known that the
exponential decay does not hold at all times: at short time we have
a Zeno-type behavior and at very long times an inverse power
fall-off. Our aim is to investigate the way the exponential decay
law appears in such systems and the characteristic time-scales for
its validity. Moreover, we would like to identify any conditions
that lead to significant divergence from the exponential decay law.

For this purpose, before we explicitly construct the probability
distribution (\ref{probab}) for this class of systems, we provide a
simple argument (using the results of section 3) that the
exponential behavior is generic\footnote{This is a variation of the
classic argument that Gamow \cite{Gam} and   Gurney with Condon
\cite{LoGu} put forward in their explanations of the alpha decay.}.
We identify the main physical assumption underlying this argument,
and we then examine whether it is consistent with the results
arising from the evaluation of $p(t)$ in the present context.

We consider the potential described earlier with $V(r) = 0$ in the
region I ($ r \in [0, a)$). For a wave-function with momentum $k_0$
with momentum spread $\sigma$, it was shown in \cite{AnSav07a} that
the probability of detection at distance L  at time $t$ is given by
\begin{eqnarray}
p_0(t)  \sim  \sqrt{\frac{1}{1 + 4 t^2 \sigma^4/M^2}}  \exp \left\{
- \frac{2 k_0^2 \sigma^2/M^2}{1 + 4 t^2 \sigma^4/M^2} \left[ t -
\frac{M(  L + \lambda_{k_0})}{k_0}\right]^2 \right\},
\end{eqnarray}
where

\begin{eqnarray} \lambda_{k_0} = \frac{M}{k_0} Im \left(
\frac{\partial \log T_k}{\partial k} \right)_{k = k_0},
\end{eqnarray}
with $T_k$ the transmission amplitude of the potential for energy
$\frac{k^2}{2M}$. (We ignored a small term in the exponential that
corresponds to the center of the initial wave-function.)

We assume that the initial state is localized in region I.
 Let $|R_{k_0}|^2$ be the reflection probability on the barrier. A
fraction $|R_{k_0}|^2$ of the particles in the ensemble will be
reflected. Since the particle is located in a bounded region (it is
reflected at $r = 0$), it will attempt to cross the barrier: in a
semi-classical approximation, this attempt will take place after
time $T = \frac{2M a}{k_0}$, and the fraction of the particles that
succeed will provide a contribution $p_1$ to the probability of
detection
\begin{eqnarray}
p_1(t)  \sim  |R_{k_0}|^2 \sqrt{\frac{1}{1 + 4 t^2 \sigma^4/M^2}}
\exp \left\{ - \frac{2 k_0^2 \sigma^2/M^2}{1 + 4 t^2 \sigma^4/M^2}
\left[ t - \frac{M(  L + \lambda_{k_0}+ 2a)}{k_0}\right]^2 \right\},
\end{eqnarray}
where the coefficient of proportionality is the same for $p_1$ and
$p_0$. Following the same argument for the multiply reflected
particle, we obtain the total probability of detection
\begin{eqnarray}
p(t) \sim \sum_{n=0}^{\infty} |R_{k_0}|^{2n} \sqrt{\frac{1}{1 + 4
t^2 \sigma^4/M^2}} \exp \left\{ - \frac{2 k_0^2 \sigma^2/M^2}{1 + 4
t^2 \sigma^4/M^2} \left[ t - \frac{M(  L + \lambda_{k_0}+ 2n
a)}{k_0}\right]^2 \right\}.
\end{eqnarray}
For times $t$ such that $t \sigma^2/M <<1$, the expression above
simplifies
\begin{eqnarray}
p(t) \sim \sum_{n=0}^{\infty} |R_{k_0}|^{2n}  \exp \left\{ - \frac{2
k_0^2 \sigma^2}{M^2} \left[ t - \frac{M(  L + \lambda_{k_0}+ 2n
a)}{k_0}\right]^2 \right\}.  \label{interm}
\end{eqnarray}

The behavior of this function is the following: until the time of
first detection $ t_0 = \frac{M(  L + \lambda_{k_0})}{k_0}$, $p(t)$
 is practically zero; then it exhibits successive sharp peaks of
width $M/(\sigma k_0)$ separated by a time interval $2 Ma/k_0$. The
height of the $n$-th peak is smaller than the height of the
$(n-1)$-th peak by a factor of $|R_{k_0}|^2$.  In effect, at time $t
= t_0 + 2n Ma/k_0$ ($n$ integer) the height of the peak will be
proportional to $ |R_{k_0}|^{2n}$. If the resolution of our time
measurements is
 coarser than the
interval $2Ma/k_0$ between successive peaks,  we can effectively
substitute the probability by the curve connecting the peaks. Hence,
for $t > t_0$
\begin{eqnarray}
p(t) \sim |R_{k_0}|^{2\frac{k_0(t - t_0)}{2Ma}} = e^{- \left|\log
|R_{k_0}|^2\right| \frac{k_0}{2Ma}(t-t_0)}.
\end{eqnarray}
We therefore have an exponential decay with decay coefficient
$\Gamma$
\begin{eqnarray}
\Gamma =  \frac{k_0}{2Ma} \log \left||R_{k_0}|^2 \right|.
\end{eqnarray}

Note that $|R_{k_0}|^2 = 1 - |T_{k_0}|^2$, where $|T_{k_0}|^2$ is
the transmission probability; if $|T_{k_0}|^2 << 1$ then $\left|\log
|R_{k_0}|^2 \right| \simeq |T_{k_0}|^2 $ and $ \Gamma =
\frac{k_0}{2Ma} |T_{k_0}|^2$. However, as time increases and $t
\sigma^2/M$ becomes of order $1$, there are no clear peaks in $p(t)$
any more and the approximation by the exponential slowly worsens.

Clearly, the description above is oversimplified. Its key assumption
is that the successive attempts of the particle to cross the barrier
are statistically independent, i.e. the $n$-th attempt has no memory
of the $(n-1)$-th attempt: the probability densities $p_n$ can then
be added. This is essentially an assumption of Markovian behavior,
which is a fundamental property of the exponential decay law.
 However, in quantum theory one does not add probabilities, but
{\em amplitudes}. Hence, one expects that in the most general case,
there will be interference terms between the different attempts of
the particle to cross the barrier: these may spoil the exponential
decay law. A full treatment should focus on the size and
contribution of these off-diagonal terms to the total probability.
We next proceed to do this.

\section{Evaluating the decay probability}

We assume a potential $V(r)$ as described in Sec. 3, for $ r \in [0,
\infty)$: $V(r) = 0$ in the regions $[0, a]$ and $[b, \infty)$.

 It is convenient to express the
mode solutions $u_k$ of Schr\"odinger's equation
\begin{eqnarray}
-\frac{1}{2M} \partial^2_r u(r) + V(r) u(r) = \frac{k^2}{2M} u(r).
\end{eqnarray}

in terms of the reflection and transmission amplitudes of the same
potential, which are defined when the variable $r$ extends from $-
\infty$ to $\infty$. Let $T_k$ and $R_k$ be the transmission and
reflection amplitude for particles coming from the left and $T_k'$
and $R'_k$ the same amplitudes for particles coming from the right.
These coefficients satisfy the conditions
\begin{eqnarray}
T_k = T_k', \hspace{2cm} |R_k| = |R_k'|, \hspace{2cm} T_k \bar{R}_k
= \bar{T}_k R'_k, \label{properties}
\end{eqnarray}
which arise from the property of the Schr\"odinger operator that the
Wronskian of two eigenfunctions with the same energy must be a
constant functions.

 The mode functions $u_k(r)$ corresponding to Dirichlet
boundary conditions at $r = 0$ are then obtained (using a multiple
scattering method)
\begin{eqnarray}
u_k(r) = \left\{ \begin{array}{c} -2i \frac{T_k}{1 + R_k}
\sin kr \; \;\; \; \; \;  \mbox{in reg. I} \\
e^{-ikr} +(R'_k - \frac{T_k^2}{1 + R_k}) e^{ikr} \; \; \;\mbox{in
reg. III} \end{array} \right.
\end{eqnarray}

Using Eqs. (\ref{properties}), we obtain  $|R'_k - \frac{T_k^2}{1 +
R_k}| = 1$. Hence, we can write the mode function $u_k(r)$ in region
III as $ e^{-ikr} - e^{i \Theta_k} e^{ikr}$, where
\begin{eqnarray}
e^{i\Theta_k} := - (R'_k - \frac{T_k^2 }{1 + R_k }). \label{theta}
\end{eqnarray}
 Imposing the Dirichlet boundary condition at $r = L$ yields the
eigenvalue equation
\begin{eqnarray}
e^{2ikL + \Theta_k} = 1,
\end{eqnarray}
with solutions $k_n$ that satisfy the set of algebraic equations
\begin{eqnarray}
k_n = \frac{n \pi}{L} - \frac{\Theta_{k_n}}{2L}, \label{eigen}
\end{eqnarray}
for all integers $n$ that lead to positive value of $k_n$.
 For $V(r) = 0$, we obtain  $k_n = n \pi/L, n = 1, 2, \ldots$.

The eigenstates $u_{k_n}(x)$ of the Hamiltonian with Dirichlet
boundary conditions are then
\begin{eqnarray}
u_{k_n}(x) = A_{k_n} \times \left\{ \begin{array}{c} -2i
\frac{T_{k_n}
}{1 + R_{k_n} } \sin k_nr \; \;\; \; \; \;  \mbox{in reg. I}\\
2i e^{-ik_nL} \sin k_n(L-r) \; \;\; \; \; \;  \mbox{in reg.
III}\end{array} \right.,
\end{eqnarray}
where $A_{k_n}$ is a normalization factor, chosen so that $\int_0^L
dx u^*_{k_n}(x) u_{k_m}(x) = \delta_{mn}$. We immediately read the
coefficient
\begin{eqnarray}
D_{k_n} = 2i A_{k_n} e^{-ik_nL} = 2i A_{k_n} e^{i\Theta_{k_n}/2}.
\end{eqnarray}
We now  choose an initial state: it should be concentrated in region
I (i.e. we assume that no element of the ensemble has decayed at $t
= 0$) and it should have mean energy $E = k_0^2/2M$, such that $k_0
>> \sigma$, where $\sigma$ is the momentum spread. For ease of
calculation, we employ a Gaussian
\begin{eqnarray}
\psi_0(r) = \frac{1}{(2 \pi \delta^2)^{1/4}} e^{ \frac{(r -
\frac{a}{2})}{4 \delta^2} + i k_0 r}. \label{initial}
\end{eqnarray}
If $e^{ - \frac{a^2}{16 \delta^2}} << 1$ then this respects with
good approximation the Dirichlet boundary condition at $r = 0$. We
could have used a different state (exactly vanishing at $r = 0$),
but it turns out the  precise form of the state makes only
difference to fine details of the probability distribution and not
to the basic features of the decay process\footnote{We have tried
different initial states with the same values for $k_0$ and $\sigma$
 and the basic features of the decay remain unchanged.}.

We then compute (with an error of order $e^{-k_0^2/\sigma^2}$)
\begin{eqnarray}
c_{k_n} = - \bar{A}_{k_n} \frac{\bar{T}_{k_n}}{1 + \bar{R}_{k_n} }
\left(\frac{4 \pi}{\sigma^2}\right)^{1/4} e^{ - \frac{(k_n -
k_0)^2}{2 \sigma^2} - i (k - k_0)a/2},
\end{eqnarray}
where $\sigma = 1/(\sqrt{2} \delta)$. The overbar denotes complex
conjugation.

Hence, the probability of detection is given by $p(t) = |z(t)|^2$,
where
\begin{eqnarray}
z(t) = - \left(\frac{4 \pi}{\sigma^2}\right)^{1/4} \sum_n
\sqrt{\frac{k_n}{4M}} 2i |A_{k_n}|^2 e^{i \Theta_k/2}
\frac{\bar{T}_{k_n} }{1 +
\bar{R}_{k_n}} \nonumber \\
\times e^{ - \frac{(k_n - k_0)^2}{2 \sigma^2} - i (k_n - k_0)a/2 - i
\frac{k_n^2t}{2M}}.
\end{eqnarray}
Using the definition (\ref{theta}) and the identities
(\ref{properties}) we find that
\begin{eqnarray}
e^{i\Theta_k} \frac{\bar{T}_{k_n}}{1 + \bar{R}_{k_n}} = \frac{T_k}{1
+ R_k }.
\end{eqnarray}
Hence,
\begin{eqnarray}
z(t) = -2i  \frac{\pi^{1/4}}{\sqrt{2M \sigma}} \sum_n |A_{k_n}|^2
e^{ik_nL}
\sqrt{k_n} \frac{T_{k_n}}{1 + R_{k_n}} \nonumber \\
\times  e^{ - \frac{(k_n - k_0)^2}{2 \sigma^2} - i (k_n - k_0)a/2 -
i \frac{k_n^2t}{2M}}. \label{zz}
\end{eqnarray}

The distance $\delta k_n$ between two neighboring eigenvalues $k_n$
and $k_{n-1}$ satisfies the equation
\begin{eqnarray}
\delta k_n = \frac{\pi}{L} - \frac{\Theta_{k_n} -
\Theta_{k_{n-1}}}{2L} \simeq \frac{\pi}{L} + \frac{\Theta'_{k_n}
\delta k_n}{2L},
\end{eqnarray}
where $\Theta_k'$ is the derivative of $\Theta_k$. We then obtain
$\delta k_n = \frac{\pi/L}{1 - \Theta'_{k_n}/2L}$. Since $L$ is a
large (macroscopic) distance, and the second order in the
denominator contributes to $\delta k_n$ terms of higher order to
$1/L$, we have effectively $\delta k_n \simeq \frac{\pi}{L}$, which
goes to zero as $L \rightarrow \infty$. For large $L$, we can
effectively substitute the summation in (\ref{zz}) with an integral.

Within the same approximation, we note the (overwhelmingly) dominant
contribution to the normalization coefficients $A_k$ comes from
region III. Hence
\begin{eqnarray}
|A_k| \simeq \frac{1}{\sqrt{2L}}.
\end{eqnarray}

Hence, we obtain the following expression for $z(t)$
\begin{eqnarray}
z(t) = \frac{-i }{\pi^{3/4} \sqrt{2 M \sigma}}
\int_{-\infty}^{\infty} dk e^{ikL} \sqrt{k} \frac{T_{k}}{1 + R_{k}}
e^{ - \frac{(k - k_0)^2}{2 \sigma^2} - i (k - k_0)a/2 - i
\frac{k^2t}{2M}}. \label{zint}
\end{eqnarray}
Noting that $(1 + R_{k})^{-1} = \sum_{n = 0}^{\infty} (-R_k)^n $, we
expand the corresponding terms around $k = k_0$ and keep only the
leading terms. Thus, we write
\begin{eqnarray}
\sqrt{k} T_k R_k^n  \simeq \sqrt{k_0} T_{k_0} R_{k_0}^n e^{i
(\lambda_{k_0} + n \beta_{k_0})(k - k_0) + (\xi_{k_0} + n s_{k_0})(k
- k_0)}, \label{expansion}
\end{eqnarray}
where we defined
\begin{eqnarray}
\lambda_{k_0} :=  \mbox{Im} \left( \frac{\partial \log T_k}{\partial
k}
\right)_{k = k_0}\label{lambdak} \\
\xi_{k_0} := \frac{1}{2k_0} + \mbox{Re} \left( \frac{\partial \log
T_k}{\partial k} \right)_{k = k_0} \label{rk}\\
\beta_{k_0} :=  \mbox{Im}  \left( \frac{\partial \log R_k}{\partial
k}
\right)_{k = k_0} \label{betak}\\
s_{k_0} := \mbox{Re}  \left( \frac{\partial \log R_k}{\partial k}
\right)_{k = k_0}. \label{sk}
\end{eqnarray}

Then $z(t)$ takes the form
\begin{eqnarray}
z(t) = \frac{-i}{\pi^{3/4}} \sqrt{\frac{k_0}{2 M \sigma}}
\sum_{n=0}^{\infty} T_{k_0} (-R_{k_0})^n \int_{-\infty}^{\infty} dk
e^{ikL} e^{i (\lambda_{k_0} + n \beta_{k_0})(k - k_0) + (\xi_{k_0} +
n s_{k_0})(k -
k_0)} \nonumber \\
\times e^{ - \frac{(k - k_0)^2}{2 \sigma^2} - i (k - k_0)a/2 - i
\frac{k^2t}{2M}} \; \; \; \; \; \; \; \; \;
\end{eqnarray}
Evaluating the Gaussian integral we arrive at the following
expression
\begin{eqnarray}
z(t) = - \frac{i}{\pi^{1/4}} T_{k_0} \sqrt{\frac{ k_0}{M \sigma
(1/\sigma^2 + i t/M)}} e^{i k_0L - i \frac{k_0^2 t}{2M}} \; \; \; \;
\; \; \; \; \; \; \; \; \; \; \; \;\; \; \; \; \; \; \; \; \; \; \;
\; \; \; \; \; \; \; \; \; \;
\; \; \;\nonumber \\
\times \sum_{n =0}^{\infty} (-R_{k_0})^n \exp \left[ \frac{ [
(\xi_{k_0} + n s_{k_0}) + i (L - a/2 + \lambda_{k_0} + n \beta_{k_0}
- k_0t/M)]^2}{2 (1/\sigma^2 + i t/M)} \right]. \; \;
\end{eqnarray}

The probability of detection is then
\begin{eqnarray}
p(t) = \frac{1}{\sqrt{\pi}}\frac{ k_0}{M \sigma \sqrt{1/\sigma^4 +
t^2/M^2}} |T_{k_0}|^2 \sum_{n = 0}^{\infty} \sum_{m =0}^{\infty}
(-1)^{n+m} R_{k_0}^n \bar{R}_{k_0}^m A_{nm}(t), \label{detprobab}
\end{eqnarray}
where
\begin{eqnarray}
A_{nm}(t) = \exp \left[ \frac{ [ (\xi_{k_0} + n s_{k_0}) + i (L -
a/2 + \lambda_{k_0} + n \beta_{k_0} - k_0t/M)]^2}{2 (1/\sigma^2 + i
t/M)} \right. \nonumber \\
\left. + \frac{ [ (\xi_{k_0} + m s_{k_0}) - i (L - a/2 +
\lambda_{k_0} + m \beta_{k_0} - k_0t/M)]^2}{2 (1/\sigma^2 - i t/M)}
\right]
\end{eqnarray}

\section{The regime for exponential decay}

\subsection{Derivation}

 Eq. (\ref{detprobab}) was obtained for a
physically reasonable choice of initial state with the additional
assumption that its momentum spread is small enough so that the
expansion (\ref{expansion}) provides a good approximation. Recalling
the discussion of Sec. 4.1, we see that the probability density
(\ref{detprobab}) exhibits  interference between the different
attempts of the particle to cross the barrier (labeled by $n$ and
$m$), and for this reason it cannot give rise to an exponential
decay law\footnote{Note that in the context of probabilities for
time in quantum theory, quantum coherence is essentially identical
with memory effects.}.

We next identify the regime in which Eq. (\ref{detprobab}) leads to
exponential decay. We expect that exponential decay arises only for
times such that $t \sigma^2/M <<1$, because at later times the peaks
in the detection probability start deteriorating and the asymptotic
long-time behavior sets-in. Note that in the present model we
assumed that $V(r) = 0$ in region I: the more physical case of a
strongly attracting potential would lead to a substantially smaller
increase of the particle's wave function spread for the time it
spends in region I, and the relevant time scale could be
significantly larger than $M/\sigma^2$.

Assuming that $t << M/\sigma^2$
\begin{eqnarray}
A_{nm}(t) = \exp \left\{ - \sigma^2 [L  + \lambda_{k_0} +
\frac{n+m}{2} \beta_{k_0} - k_0t/M]^2  \right. \nonumber \\
\left.
 + \sigma^2 (\xi_{k_0} + \frac{n+m}{2} s_{k_0})^2
- \frac{\sigma^2}{4} (n - m)^2 (\beta_{k_0}^2 - s_{k_0}^2) \right.
\nonumber
\\
\left. + i \sigma^2 (n - m) [\xi_{k_0} \beta_{k_0} + s_{k_0} (L +
\lambda_{k_0} - k_0 t/M) + (n+m) s_{k_0} \beta_{k_0}] \right\}.
\label{Anm}
\end{eqnarray}
In the expressions above, we substituted  $L - a/2$ (the distance of
the detector from the center of the initial state) with $L$, since
$L>> a$. We see that $A_{nm}(t)$  has strong peaks at times $t =
\frac{M(L  + \lambda_{k_0} + \frac{n+m}{2} \beta_{k_0})}{k_0}$, i.e.
at multiples of $\beta_{k_0}$ after the time $t_0 = M(L  +
\lambda_{k_0})/k_0$ of first detection\footnote{As shown in
\cite{AnSav07a}, $\frac{M \lambda_{k_0}}{k_0}$ is the delay time for
the recording of the particles, which is due to the presence of the
barrier.}. There are also oscillating terms proportional to $(n-m)$,
hence there are no sharp instants of detection. However, large
differences in value between $n$ and $m$ are suppressed by a term
$e^{ - \sigma^2 (n-m)^2 (\beta_{k_0}^2 - s_{k_0}^2)}$. Hence, if
\begin{eqnarray}
e^{\sigma^2 (\beta_{k_0}^2 - s_{k_0}^2)} >>1 \label{condition1}
\end{eqnarray}
the values of $A_{nm}$ for $n \neq m$ are very small for all times
$t$. Effectively,
\begin{eqnarray}
A_{nm}(t) = \delta_{nm} \; e^{ - \sigma^2 [L  + \lambda_{k_0} + n
\beta - k_0t/M]^2 + \sigma^2 (\xi_{k_0} + ns_{k_0})^2 }.
\end{eqnarray}
Substituting into Eq. (\ref{detprobab}), we obtain
\begin{eqnarray}
p(t) = \frac{k_0 \sigma}{\sqrt{\pi} M} |T_{k_0}|^2 \sum_{n =
0}^{\infty} |R_{k_0}|^{2n} e^{ - \sigma^2 [L  + \lambda_{k_0} + n
\beta - k_0t/M]^2 + \sigma^2 (\xi_{k_0} + ns_{k_0})^2 }. \label{pp2}
\end{eqnarray}
This probability distribution behaves as follows. For $t < t_0$,
$p(t) \simeq 0$. At $t = t_0$ there is a peak of width
$\frac{M}{2k_0 \sigma}$ (corresponding to first detection), and then
there are successive peaks of the same width centered around $t_n =
t_0 + n \frac{M \beta_{k_0}}{k_0}$ with an amplitude differing by a
factor of $|R_{k_0}|^{2n} e^{\sigma^2 (\xi_{k_0} + ns_{k_0})^2}$.

This is similar to the semi-classical description of Sec. 3, namely
that the peak at time $t_n$ corresponds to the (n+1)-th attempt of
the particle to cross the barrier. The interpretation of the
quantity $\beta_{k_0}$ is simpler if we assume that the potential
$V(r)$ is parity-symmetric, or more precisely if $V'(x) := V(x - a -
\frac{d}{2})$ for $d = b - a$ satisfies $V'(x) = V'(-x)$. For
$V'(x)$, $R_k = R'_k$ and  Eqs. (\ref{properties}) imply that
\begin{eqnarray}
\mbox{Arg} R_{k} = - \frac{\pi}{2} + \mbox{Arg} T_k + 2ka + kd,
\end{eqnarray}
so that $\beta_{k_0} = \lambda_{k_0} + 2a + d$. Hence, the distance
between two successive peaks in $p(t)$ equals
\begin{eqnarray}
\frac{M \beta_{k_0}}{k_0} = t_{cross} + \frac{2Ma}{k_0},
\end{eqnarray}
where $t_{cross} = \frac{M}{k_0} \left( d + \left(\frac{\partial
T_k}{\partial k} \right)_{k = k_0} \right)$ is the time it takes a
particle to cross the barrier region. Hence, $M \beta_{k_0}/k_0$ is
the sum of the time it takes a classical free particle between two
attempts to cross the barrier plus the time for crossing the
barrier.

 However, Eq. (\ref{pp2}) does not correspond to exponential decay:
the ratio of the amplitude of two successive peaks is not constant
due to the presence of the term $e^{\sigma^2 (\xi_{k_0} +
ns_{k_0})^2}$. We assume that the contribution of this term is much
smaller than that of $|R_{k_0}|^{2n}$, i.e. we consider values of
$n$ such that
\begin{eqnarray}
\sigma^2 (\xi_{k_0} + ns_{k_0})^2 << n \log|R_{k_0}|^2.
\label{condition2}
\end{eqnarray}

In this regime,
\begin{eqnarray}
p(t) = \frac{ k_0 \sigma}{\sqrt{\pi} M} |T_{k_0}|^2 \sum_{n =
0}^{\infty} |R_{k_0}|^{2n} e^{ - \sigma^2 [L  + \lambda_{k_0} + n
\beta_{k_0} - k_0t/M]^2 }. \label{prob-1}
\end{eqnarray}
The arguments of Sec. 4.1 can now be used in a straightforward
manner. If the temporal resolution of measurements is larger than
the distance between successive peaks, then for $t > t_0$, we can
substitute $p(t)$ by the curve connecting these peaks. However, we
have to preserve the normalization. The probability corresponding to
each Gaussian in the sum (which equals $\sqrt{\pi}M \sigma/k_0$)
must be spread within an interval of width $M \beta_{k_0}/k_0$ (i.e.
the distance between two peaks.) This yields (for $t > t_0$)
\begin{eqnarray}
p(t) \simeq \frac{ k_0 }{M \beta_{k_0}} |T_{k_0}|^2
\left(|R_{k_0}|^2\right)^{k_0\frac{t - t_0}{M \beta_{k_0}}} = \frac{
k_0}{M \beta_{k_0}} |T_{k_0}|^2 e^{ - \left|\log |R_{k_0}^2| \right|
\frac{ k_0}{M \beta_{k_0}} (t - t_0)}. \label{decay2}
\end{eqnarray}
Since we assumed $|T_{k_0}|^2 << 1$, the expression above becomes
\begin{eqnarray}
p(t) =  \Gamma e^{- \Gamma (t - t_0)} = - \frac{d}{dt} e^{- \Gamma(t
-t_0)}, \label{decay}
\end{eqnarray}
where $\Gamma$ is the decay rate
\begin{eqnarray}
\Gamma = \frac{ k_0}{M \beta_{k_0}} |T_{k_0}|^2. \label{decay_rate}
\end{eqnarray}

We thus obtained the standard expression for the exponential decay
law.

 Note that $t_d = M \beta_{k_0}/k_0$ is the distance between
two successive peaks and that $\Gamma t_d = |T_{k_0}|^2 << 1$. The
characteristic time-scale $\Gamma^{-1}$ associated to the decay must
be much larger than the distance between successive peaks for the
substitution of (\ref{prob-1}) with (\ref{decay2}) to make sense:
monitoring the detection at time scales of order $\Gamma^{-1}$
should not allow one to distinguish the fine structure of the
probability distribution (\ref{prob-1}).

\begin{figure}[htp]
\centering
\includegraphics*[0pt, 350pt][500pt, 550 pt]{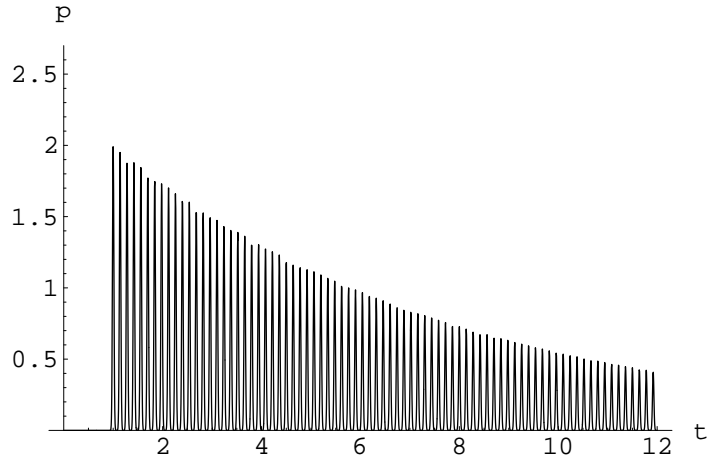}
\caption{ \small The probability distribution of the time-of-arrival
in the exponential decay regime. The exponential curve is the
envelope of the distribution's curve, and it has non-overlapping
successive peaks at a scale of $M \beta_{k_0}/k_0$.}
\end{figure}

\subsection{The conditions for exponential decay}

We now examine in more detail the conditions that are necessary for
the derivation of the exponential decay law. Most important amongst
them is the fact that the transmission probability $|T_{k_0}|^2$
must be much smaller than unity. Otherwise, there is no regime for
exponential decay. This is the case for example, if the particle's
energy is close to the peak of the potential barrier. In effect,
this condition involves a separation of time-scales: the typical
decay time should be much larger than the characteristic time
between two different attempts of the particle to cross the barrier.
In open quantum systems, this separation of time-scales is a
necessary condition for Markovian behavior--essentially implying
that the memory effects can be ignored in the derivation of the
evolution law. The context is different here, but the principle
remains the same.

The condition (\ref{condition2})  provides an upper time-limit for
the  validity of  exponential decay. We first note that the
definitions (\ref{rk}) and (\ref{sk}) together with the identities
(\ref{properties}) we obtain
\begin{eqnarray}
s_{k_0} = - (\xi_{k_0} - \frac{1}{2k_0})
\frac{|T_{k_0}|^2}{|R_{k_0}|^2}. \label{ident1}
\end{eqnarray}
Since by (\ref{rk}) $ \xi_{k_0} > \frac{1}{2k_0}$ ($|T_k|$ increases
with $k$)
\begin{eqnarray}
\left| \frac{s_{k_0}}{\xi_{k_0}} \right| <
\frac{|T_{k_0}|^2}{|R_{k_0}|^2}.
\end{eqnarray}
Hence, for $n$ less or of the order of $|\xi_{k_0}/s_{k_0}|$, the
left-hand-side of the inequality is $n$-independent and of order
$\sigma^2 \xi_{k_0}^2$. Hence, the factor $e^{\sigma^2 (\xi_{k_0} +
ns_{k_0})^2}$ does not affect significantly the exponential decay
law--at most by a multiplicative factor $e^{\sigma^2 r^2_{k_0}}$.
However, for larger values of $n$ (time increasing) the  term $n^2
\sigma^2 s_{k_0}^2$ dominates the left-hand side and the condition
(\ref{condition2}) becomes $n << \frac{|T_{k_0}|^2}{\sigma^2
s_{k_0}^2}$, or in terms of time $t$
\begin{eqnarray}
t - t_0 << \frac{M \beta_{k_0} |T_{k_0}|^2}{k_0 \sigma^2 s_{k_0}^2}.
\end{eqnarray}

Substituting $s_{k_0}$ from Eq. (\ref{ident1}) and using Eq.
(\ref{decay_rate}) we obtain (for $|T_{k_0}|^2 << 1$)
\begin{eqnarray}
\Gamma (t - t_0) << \left[ \sigma^2 w_{k_0}^2 \right]^{-1}.
\label{decaylimit}
\end{eqnarray}
where we wrote $ w_{k_0} = \left(\frac{\partial \log |T_k|}{\partial
k}\right)_{k = k_0}$.

Hence, for the exponential decay law to be valid for times much
larger than the decay time $\Gamma^{-1}$, it is necessary that
\begin{eqnarray}
\sigma \left(\frac{\partial \log |T_k|}{\partial k}\right)_{k = k_0}
<< 1, \label{cond3}
\end{eqnarray}
or, defining the variation in  transmission probability according to
the momentum spread $\sigma$ of the initial state $\Delta |T_{k_0}|
: = \sigma \left(\frac{\partial |T_k|}{\partial k}\right)_{k =
k_0}$,
\begin{eqnarray}
\Delta |T_{k_0}| << |T_{k_0}|.
\end{eqnarray}
In effect, the interference between decay channels characterized by
different decay coefficient must be negligible, in order to obtain
an exponential decay law. Moreover, the time-scale $\Gamma^{-1}
|T_{k_0}|^2/(\Delta|T_{k_0}|)^2$ characterizes the breakdown of the
exponential decay law.

We finally examine condition (\ref{condition1}). Since, for
$|T_{k_0}| << 1$, $|s_{k_0}| \simeq |T_{k_0}|^2 |w_{k_0}|$,
(\ref{cond3}) implies that the condition (\ref{condition1}) can only
be satisfied if
\begin{eqnarray}
\sigma \beta_{k_0} >>1. \label{condition4}
\end{eqnarray}
This means that  the position spread $\sigma^{-1}$ of the initial
state must be substantially smaller than the effective distance
traversed by the particle between an unsuccessful attempt to cross
the barrier and a successful one. If this does not hold, then there
is quantum interference between the different attempts and the
off-diagonal elements of $A_{mn}(t)$ will not be suppressed. An
initial state with a spread equal to $a$ can only give rise to an
exponential decay law if the effective tunneling length is very
large.

We see therefore that the exponential decay law involves a specific
intermediate regime for the characteristic features of the initial
state. The position spread must be sufficiently small so that there
is no interference between different attempts to cross the barrier,
but it cannot be too small because the variation $\Delta |T_{k_0}|$
will become too large and create interference of decays with
different characteristic time-scales.

We also note that the conditions (\ref{cond3}) and
(\ref{condition4}) can be satisfied simultaneously only if
\begin{eqnarray}
\mbox{Im}  \left( \frac{\partial \log R_k}{\partial k} \right)_{k =
k_0} >> \mbox{Re}  \left( \frac{\partial \log T_k}{\partial k}
\right)_{k = k_0}.
\end{eqnarray}
This is a necessary condition that any potential has to satisfy {\em
irrespective of the choice of the initial state}, if the decay
probability is to be characterized by an exponential regime.

To summarize, there are three conditions on the potential and on the
initial state that have to be satisfied, for a regime of an
exponential decay to exist.
\\
1. $|T_{k_0}|^2 << 1$. This is necessary for the clear
identification of the decay behavior over the fine structure
exhibited by $p(t)$ at the time scale of $M \beta_{k_0}/k_0$.\\
2. $\sigma \beta_{k_0} >> 1$. This is sufficient for the suppression
of interferences between different crossing attempts and the
validity of the semiclassical picture of Sec. 4.1. \\
3. $\sigma w_{k_0} << 1$. This guarantees that there is no
interference between processes characterized by substantially
different decay coefficients.
\\ \\
We also note that Eq. (\ref{detprobab}), to which the conditions 1-3
above refer, is obtained by keeping only the first order in the
expansion for the logarithm of the reflection and transmission
amplitudes. We have therefore assumed the following conditions
\begin{eqnarray}
\sigma \mbox{Re} \left( \frac{\partial^2 \log T_k}{\partial k^2}
\right)_{k = k_0}  <<   w_{k_0} \label{extra1} \\
\sigma \mbox{Im} \left( \frac{\partial^2 \log T_k}{\partial k^2}
\right)_{k = k_0}  <<   \lambda_{k_0} \\
\sigma  \mbox{Re} \left( \frac{\partial^2 \log R_k}{\partial k^2}
\right)_{k = k_0}  << s_{k_0} \\
\sigma  \mbox{Im} \left( \frac{\partial^2 \log R_k}{\partial k^2}
\right)_{k = k_0}  << \beta_{k_0}. \label{extra4}
\end{eqnarray}

\subsection{Special cases}

We now examine the regime for exponential decay for the case of a
square potential barrier, i.e. if $V(r) = V_0$ for $ x \in [a, b]$
and $V(r) = 0$ otherwise.
 Defining
$\gamma_k = \sqrt{2MV_0 - k^2}$, we obtain the following values for
the coefficients $T_k, R_k$
\begin{eqnarray}
T_k = \frac{2 k}{\gamma_k} e^{-ikd} \frac{2k \gamma_k [2 k \gamma_k
\cosh \gamma_k d - i  (\gamma^2_k - k^2) \sinh \gamma_k d]}
{4 k^2 \gamma_k^2 + (\gamma_k^2 + k^2) \sinh^2 \gamma_k d} \\
R_k = -i e^{2ika }  \frac{(\gamma^2 + k^2) [2 k \gamma \cosh
\gamma_k d - i  (\gamma_k^2 - k^2) \sinh \gamma_k d]}{4 k^2
\gamma_k^2 + (\gamma_k^2 + k^2) \sinh^2 \gamma_k d} .
\end{eqnarray}

There are two limits, in which the expressions above simplify.
\\ \\
1. The limit of a long barrier $\gamma_k d >> 1$, for which

\begin{eqnarray}
T_k &\simeq& e^{-ikd} e^{-\gamma_k d} \frac{4 k
\gamma_k}{(\gamma_k^2 + k^2)^2} [2k \gamma_k - i (\gamma_k^2 - k^2)]
\label{tklong}
\\
R_k &\simeq& e^{2ika} \frac{-(\gamma_k^2 - k^2) + ik\gamma_k}{4
\gamma_k^2 } \label{rklong}
\end{eqnarray}

At this limit, $|T_{k_0}| << 1$. We find,
\begin{eqnarray}
\beta_{k_0} = 2 (a + \gamma_{k_0}^{-1}).
\end{eqnarray}
The condition $\sigma \beta_{k_0} >> 1$ follows by the assumption
that $\sigma a >> 1$.

We also obtain
\begin{eqnarray}
w_{k_0} \simeq k_0^{-1} + 2 d \frac{k_0}{\gamma_{k_0}}.
\end{eqnarray}
Since $\sigma/k_0 << 1$,  the condition $ \sigma w_{k_0} << 1$ can
only be satisfied if $\sigma d k_0/\gamma_{k_0} << 1$. If $d$ is of
the order of $a$ or smaller, then it is necessary that
$k_0/\gamma_{k_0} << 1$, i.e. that the particle's energy
$\frac{k_0^2}{2M}$ is much smaller than the barrier's height $V_0$.
\\ \\
2. The  limit of the delta function (very short) barrier. It is
obtained by letting $V_0 \rightarrow \infty$ and $d \rightarrow 0$
such that $V_0 d$ is a constant (we denote this constant as
$\kappa/M$). At this limit, $\gamma_k d \simeq \sqrt{\kappa d}$ and
\begin{eqnarray}
T_k = \frac{1}{1 + i \kappa/k} \label{tkdelta} \\
R_k = e^{2ika} \frac{1}{1 + i k/\kappa}. \label{rkdelta}
\end{eqnarray}

At this limit, the condition $|T_{k_0}| << 1$ is satisfied only if
$k_0/\kappa << 1$. In this regime,
\begin{eqnarray}
\beta_{k_0} = 2 a + \frac{1}{\kappa}.
\end{eqnarray}
Again, the condition $\sigma \beta_{k_0} >> 1$ follows from the
assumption that $\sigma a >> 1$. We also compute
\begin{eqnarray}
w_{k_0} = k_0^{-1}.
\end{eqnarray}
The condition $\sigma w_{k_0} << 1$ is therefore satisfied for all
values of $\kappa$.

In both of the examples above, the validity of the conditions 1-3
implies the validity of the conditions (\ref{extra1}--\ref{extra4})
on the second derivatives of the transmission and reflection
amplitudes.

\section{Beyond exponential decay}

\subsection{ The long-time limit(s)}

Eq. (\ref{decaylimit}) implies that there is an upper limit to the
duration of the exponential decay. In fact, we can find the
effective probability distribution corresponding to (\ref{pp2}) by
ignoring its fine structure without assuming the condition
(\ref{condition2}). Following the same steps as in the derivation of
Eq. (\ref{decay}) we obtain (for $|T_{k_0}| << 1$) an equation for
$p(t)$  which is valid even for times at which condition
(\ref{decaylimit}) breaks down
\begin{eqnarray}
p(t) = \Gamma \exp \left[ - \Gamma (t - t_0) + \sigma^2 w_{k_0}^2[ 1
- \Gamma (t - t_0)]^2 \right].
\end{eqnarray}
However, the equation above cannot hold for all times, because it
would not define a normalized probability. It is only valid for the
regime that characterizes the breakdown of exponential decay. It
shows an increase in the detection rate.

There is another limit forced by the eventual spread of the
particle's wave function in region I. In our model, we assumed that
$V(r) = 0$ in region I: the time scale in which the wave-packet
spread becomes significant is of the order of $t \sim M/ \sigma^2 $.
At this time-scale, the fine structure of $p(t)$ becomes blurred.
Eq. (\ref{detprobab}) leads to the following asymptotic form for
$p(t)$,

\begin{eqnarray}
p(t) = \frac{ 2 k_0 |T_{k_0}|^2}{\sigma |1 + R_{k_0} e^{2
k_0(\beta_{k_0} - i s_{k_0})}|^2} \; t^{-1}. \label{longt}
\end{eqnarray}

Eq. (\ref{longt}) may be valid for times much larger than $M/
\sigma^2$, but it is unacceptable as the limiting behavior for $t
\rightarrow \infty$: the integral $\int_0^{\infty} dt p(t)$ diverges
logarithmically, while by construction it should take a value less
than unity. The reason is easy to identify: Eq. (\ref{detprobab})
involves an expansion of the logarithm of the transmission and
reflection amplitudes, where only the first term is kept. However,
the probability at the long time limit receives contributions from
the {\em deep infrared} values of $k$, for which $T_k \simeq 0$; the
expansion of $\log T_k$ is then inadequate as $\log T_k \rightarrow
- \infty$. To find the behavior of $T_k$ at this limit we have to go
back to a prior stage of our calculation, namely to Eq. (\ref{zint})
for $z(t)$. Substituting $k = y /\sqrt{t}$ and taking the dominant
terms as $t \rightarrow \infty$, we obtain
\begin{eqnarray}
z(t) = - \frac{i}{\sqrt{\pi M \sigma}} \int_{-\infty}^{\infty}
\frac{y dy}{t^{3/2}} \frac{T_{y/\sqrt{t}}}{1 + R_{y/\sqrt{y}}} e^{-i
y^2/2M}. \label{zlong}
\end{eqnarray}
As $t \rightarrow \infty$, the dominant contribution comes from the
values of $T_k$ around $k = 0$. Let $T_k \sim k^{\alpha}$ as $k
\rightarrow 0$. Since $|R_{k=0}| = 1$, we see that Eq. (\ref{zlong})
leads to a behavior $z(t) \sim 1/t^{3/4 + \alpha/2}$. Hence, as $t
\rightarrow \infty$
\begin{eqnarray}
p(t) \sim \frac{1}{ t^{3/2 + \alpha} }.
\end{eqnarray}

For the square-well potential, $\alpha = 1$, hence $p(t)$ drops
asymptotically as $t^{-5/2}$.

\subsection{Non-exponential decays}

We remind that in the derivation of the POVM (\ref{probab}) we took
the limit that the `response time' $\tau$ of the detector is much
larger than energy spread of the initial state \cite{AnSav07a}. As a
result, there is no Zeno-like behavior at early times in the POVM
(\ref{probab}): its predictions are only valid for times much larger
than $\tau$. To study the detection probability at time scales
relevant to the Zeno effect we should employ the  probability
density (2.19) of Ref. \cite{AnSav07a} that has an explicit
$\tau$-dependence: the temporal resolution of the detector places
limits on how precisely one can identify the Zeno-like behavior.

We now examine the issue whether there exist physical systems with
no exponential decay regime. Clearly, the condition $|T_{k_0}| << 1$
is crucial. If this does not hold, the probability distribution
exhibits a fine structure due to the details of the initial state
and it is not meaningful to talk about an exponential regime. The
ensemble will decay almost fully after the first few attempts of the
particles to cross the barrier: the relevant time-scale will be
microscopic and of the order of the $M \beta_{k_0}$. Hence, the
condition $|T_{k_0}| << 1$ is necessary for a decay characterized by
a macroscopically distinguishable time-scale. If this condition does
not hold, the decay law will be highly irregular--such decays were
identified in \cite{ExFr07} for the special case of a delta-function
barrier.

A question that arises in this context, is whether it is possible to
predict the fine structure (peaks in probability) at the scale of $M
\beta_{k_0}$, say in potentials for which $\beta_{k_0}$ takes very
large values. We believe that in general it is not: the reason is
that in a realistic preparation of a decaying system we have no
control over the center of the initial wave packet. The only
conditions we can safely identify is its mean energy, the energy
mean deviation and the fact that it is localized in region I. This
means that we cannot make a reasonable prediction about the fine
structure of $p(t)$, which depends on other parameters of the state.
Moreover, our lack of control over the details of the initial state
may imply that the most adequate description may be in terms of a
mixed state: we will then obtain a convex combination of probability
distributions of the type (\ref{detprobab}),in which case the peaks
in the probability distribution will be blurred. Hence, only the
behavior of $p(t)$ at a sufficiently coarse-grained time scale can
be predicted, because it is insensitive to the details of the
initial state.

The condition $\sigma \beta_{k_0} >> 1$ essentially requires that
the wave-function's position spread is much smaller than the size
$a$ of region I. This depends on the physical system under
consideration. For example, in  alpha-decay the spread both $\sigma$
and $a$ are determined by the specifics of the interaction between
the alpha particle and the other nucleons: they are not free
parameters that may be varied by the experimentalist. However, in
condensed matter systems, it might be possible to control at least
the value of $a$. The violation of this condition implies that the
interference terms between different attempts to cross the barrier
become substantial. It should also be noted that $\sigma /\beta$ is
the ratio between the distance of two successive peaks and the width
of each peak. Hence, if this quantity is of the order of unity the
probability distribution is much smoother at the scale of
$M\beta_{k_0}/k_0$ than it is in the case we considered earlier.

We noted that for the symmetric potentials $\beta_{k_0}$ is larger
than $2a$. Hence, $\sigma \beta_{k_0}$ is in this case larger than
$2 \sigma a$. Now $1/2\sigma < a$ (since the state must be localized
in region I); hence, $\sigma \beta_{k_0}$ cannot take values lower
than unity in these systems. This implies that to first
approximation we can keep only the terms with $|m - n| = 0 , 1$ in
Eq. (\ref{Anm}). Dropping for simplicity, the $\lambda_{k_0}$ term
that only changes slightly the value of $t_0$ and taking
$|s_{k_0}|<< \beta_{k_0}$ we obtain the following expression
\begin{eqnarray}
p(t) \simeq \frac{k_0 \sigma}{\sqrt{\pi} M} |T_{k_0}|^2 \sum_{N =
0}^{\infty} |R_{k_0}|^{2N}  e^{- \sigma^2 (L + N \beta_{k_0} -
k_0t/M)^2 + \sigma^2
(\xi_{k_0} + N s_{k_0})^2} \nonumber \\
\times \sin^2 \left( \theta_{k_0} + \frac{1}{2} \xi_{k_0}
\beta_{k_0} + \frac{1}{2} s_{k_0} (L - k_0t/M + 2 N \beta_{k_0})
\right),
\end{eqnarray}
where $\theta_{k_0} = \mbox{Arg} R_{k_0}$. The distance $M
\beta_{k_0}/k_0$ between two peaks is now of the order of the peak's
width $\sigma$. At times $t_N = t_0 + N M \beta_{k_0}/k_0$, we
obtain
\begin{eqnarray}
p(t_N) \simeq  \frac{ k_0 }{M \beta_{k_0}} |T_{k_0}|^2
\left(|R_{k_0}|^{2}\right)^{ k_0(t_N-t_0)/M\beta_{k_0}} e^{ \sigma^2
(\xi_{k_0} + k_0 s_{k_0}(t_N-t_0)/M \beta_{k_0})^2} \nonumber \\
\times\left(\ldots + \frac{e^{-4\sigma^2 \beta_{k_0}^2
}}{|R_{k_0}|^4} + \frac{e^{-\sigma^2 \beta_{k_0}^2}}{|R_{k_0}|^2} +
1 + e^{- \sigma^2 \beta_{k_0}^2} |R_{k_0}|^2 + e^{-4\sigma^2
\beta_{k_0}^2}
|R_{k_0}|^4 + \ldots \right) \nonumber \\
  \nonumber \sin^2 \left( \theta_{k_0} + \frac{1}{2}
\beta_{k_0} (r_{k_0} + M s_{k_0}(t_N-t_0)/k_0) \right).
\end{eqnarray}

At times, such that $M(t_N - t_0) \sigma_{k_0}/k_0 << r_{k_0}$, and
assuming $|T_{k_0}|^2 << 1$  we obtain an exponential decay law. For
$t > t_0$
\begin{eqnarray}
p(t) \simeq \frac{ k_0 }{M \beta_{k_0}} (1 + 2 e^{-\sigma^2
\beta_{k_0}^2} )|T_{k_0}|^2 \sin^2(\theta_{k_0} + \frac{1}{2}
r_{k_0} \beta_{k_0}) e^{ - \frac{k_0 }{M \beta_{k_0}}|T_{k_0}|^2 (t
- t_0)}, \; \; \; \;
\end{eqnarray}
where we only kept the leading order terms in $e^{\sigma^2
\beta_{k_0}^2}$ and as in (\ref{decay}) we took $e^{\sigma^2
r_{k_0}^2} \simeq 1$. Thus, we see that the relaxation of the
condition $\sigma \beta_{k_0} << 1$ to $\sigma \beta_{k_0} \sim 1$
only affected the regime of exponential decay by a multiplicative
constant factor. To obtain a qualitatively different behavior, one
would have to assume $\sigma \beta_{k_0} << 1$, which is
inadmissible at least for symmetric potentials. This implies that
the suppression of the off-diagonal terms in $A_{nm}(t)$ of Eq.
(\ref{Anm}) is rather generic. Still, this result crucially depends
on the fact that the potential vanishes in region I. A strongly
attracting potential could lead to a value of $\beta_{k_0}$
substantially smaller than $2a$, whence the limit $\sigma
\beta_{k_0}<< 1$ could be applicable.

We next examine the consequences of violating the condition $\sigma
w_{k_0} << 1$. Usually, the transmission probability is a bounded
function of $k$ and for small values of $k_0$ (energy substantially
lower than the barrier's height), its rate of change is  slow
(unless the barrier varies rapidly at the scale of $k_0^{-1}$).
However, even for the simple case of a long square barrier
potential, there is a regime in which this condition is violated. If
$\gamma_{k_0} d
>>1$, $|T_{k_0}|^2 << 1$, irrespective of the value of $k_0$: hence
there is a distinguishable macroscopic decay time-scale. If the
particle's energy, is close to the barrier's height, so that
$\gamma_{k_0}/k_0$ is a very small number, it suffices that
$\sigma/\gamma_{k_0}$ is of the same order as $\gamma_{k_0}/k_0$ to
get a violation. However, in this regime we also have a violation of
the conditions (\ref{extra1}--\ref{extra4}), so Eq.
(\ref{detprobab}) is not an adequate approximation. One would have
to keep higher order terms in the expansion for the logarithm of the
reflection and transmission amplitudes. In this case, there are no
meaningful detection peaks (even ones involving interference): the
decay is going to be non-exponential. Note that in this regime,
$k_0$ is near the potential's threshold, and in this sense it is
analogous to a regime of non-exponential decay identified in
\cite{BeEk}--see also \cite{Jietal}. In general, there is going to
be non-exponential decay in any potential characterized by a regime
of energies, in which the transmission probability varies rapidly.
This does not require that the energy be close to the potential's
peak. For example, in any barrier which can be approximated by the
double-step potential
\begin{eqnarray}
V(x) = \left\{ \begin{array}{c} V_1, \; \; \; \; \; \; \; \; \; \; \; \; \; \; \; \; a \leq x \leq b \\
                                V_2 (> V_1), \;  \;  \; \; \; \; \; b < x < c
                                \end{array} \right.
\end{eqnarray}

the regime of energies around $k_0 = \sqrt{2MV_1}$ will give rise to
non-exponential decay.

To  obtain a picture for the behavior of such non-exponential decays
without moving beyond the validity of the conditions
(\ref{extra1}--\ref{extra4}), we consider an initial state which is
a superposition of two Gaussians of the same width $\sigma$, but
with different values of momenta $k_1$ and $k_2$

\begin{eqnarray}
\psi_0(r) = \frac{1}{\sqrt{2}} \left( \frac{1}{(2 \pi
\delta^2)^{1/4}} e^{ \frac{(r - \frac{a}{2})}{4 \delta^2} + i k_1 (r
- \frac{a}{2})} + \frac{1}{(2 \pi \delta^2)^{1/4}} e^{ \frac{(r -
\frac{a}{2})}{4 \delta^2} + i k_2 (r - \frac{a}{2})} \right).
\label{superpos}
\end{eqnarray}

This initial state, being a superposition of two states with
different mean energy, could be relevant for the description of
quantum beats \cite{QB}in systems decaying through tunneling--for
example \cite{qbn}: we shall see that it leads to an oscillating
behavior of the decay probability.

 We assume that $  \frac{\left||T_{k_1}| -
|T_{k_2}| \right|}{\left||T_{k_1}| + |T_{k_2}| \right|}$ is of the
order of unity. In cases such as the two step potential this does
not necessitate that the absolute value of $ q = k_1 - k_2$ is
large--it suffices that it is substantially larger than $\sigma$.
The conditions (\ref{extra1}--\ref{extra4}) then need not be
violated in the calculation. We neglect for simplicity all terms
that are usually small in the exponential phase, i.e. we employ the
same approximations involved in the derivation of Eq.
(\ref{prob-1}).  We also drop the small $\lambda_k$ term in the
exponential. Then, we obtain
\begin{eqnarray}
p(t) =  \frac{ k_1 \sigma}{2 \sqrt{\pi}M} |T_{k_1}|^2 \sum_{n =
0}^{\infty} |R_{k_1}|^{2n} e^{ - \sigma^2 [L    + n \beta_{k_1} -
k_1t/M]^2 } \hspace{3cm} \nonumber \\
+  \frac{ k_2 \sigma}{2 \sqrt{\pi}M} |T_{k_2}|^2 \sum_{n =
0}^{\infty} |R_{k_2}|^{2n} e^{ - \sigma^2 [L  + n \beta_{k_2} -
k_2t/M]^2 } \hspace{3cm}  \nonumber \\
+ \frac{ \sigma \sqrt{k_1 k_2}}{\sqrt{\pi}M} \mbox{Re} \left[
e^{iq(L - \frac{k_0}{M}t)} T_{k_1} \bar{T}_{k_2}
\sum_{n,m=0}^{\infty} (-R_{k_1})^n (-\bar{R}_{k_2})^m e^{-\sigma^2(L
+ \frac{n\beta_{k_1}
+ m \beta_{k_2}}{2} - \frac{k_0t}{M})^2} \right. \nonumber \\
\left. \times e^{ - \sigma^2 (\frac{qt}{M} - \frac{n \beta_{k_1} -
m\beta_{k_2}}{2})^2 - \frac{\sigma^2}{2} (n \beta_{k_1} -
m\beta_{k_2})^2} \right], \;\;\;\;\; \label{pppp}
\end{eqnarray}
where $k_0 = (k_1 +k_2)/2$. We assume that the coefficient $\beta_k$
does not vary much between $k_1$ and $k_2$ --which is reasonable
since the dominant contribution to $\beta_k$ is $2a$ (at least for
the symmetric potentials). Then the off-diagonal elements in the
last term are suppressed and this term becomes
\begin{eqnarray}
\frac{ \sigma \sqrt{k_1 k_2}}{\sqrt{\pi} M}
e^{-\frac{\sigma^2q^2t^2}{M^2}} \mbox{Re} \left( e^{iq(L -
\frac{k_0}{M}t)} T_{k_1} \bar{T}_{k_2}  \sum_{n=0}^{\infty} (R_{k_1}
\bar{R}_{k_2})^n e^{- \sigma^2 (L + n
\frac{\beta_{k_1}+\beta_{k_2}}{2} - \frac{k_0t}{M})^2} \right).
\end{eqnarray}
The same analysis as in Sec. 4 then yields for $t > t_0$
\begin{eqnarray}
p(t) &=& \frac{\Gamma_1}{2} e^{- \Gamma_1t} + \frac{\Gamma_2}{2}
e^{-\Gamma_2 t} \nonumber
\\ &+& \frac{2  \sqrt{k_1 k_2}}{M(\beta_{k_1}  +
\beta_{k_2})} e^{-\frac{\sigma^2q^2t^2}{M^2}} \mbox{Re} \left(
e^{-i\frac{k_0q}M(t - t_0)}T_{k_1} \bar{T}_{k_2} (R_{k_1}
\bar{R}_{k_2})^{\frac{2 k_0 (t - t_0)}{M (\beta_{k_1} +
\beta_{k_2})} }\right), \; \; \;  \; \; \; \; \; \; \;\;\;
\label{ttt}
\end{eqnarray}

where $\Gamma_{1,2} = \frac{ k_{1,2}}{M \beta_{k_{1,2}}}
|T_{k_{1,2}}|^2$.

We see then that $p(t)$ is a convex combination of two exponential
decay terms together with an interference term that becomes
negligible at times $t >> \frac{M}{q \sigma}$. Near a threshold the
decay rates $\Gamma_{1,2}$ may be substantially different even for
relatively small values of $q$. This implies that the oscillatory
behavior arising from the interference term may persist long enough
to be macroscopically distinguishable.

\begin{figure}[htp]
\centering
\includegraphics*[0pt, 350pt][500pt, 550 pt]{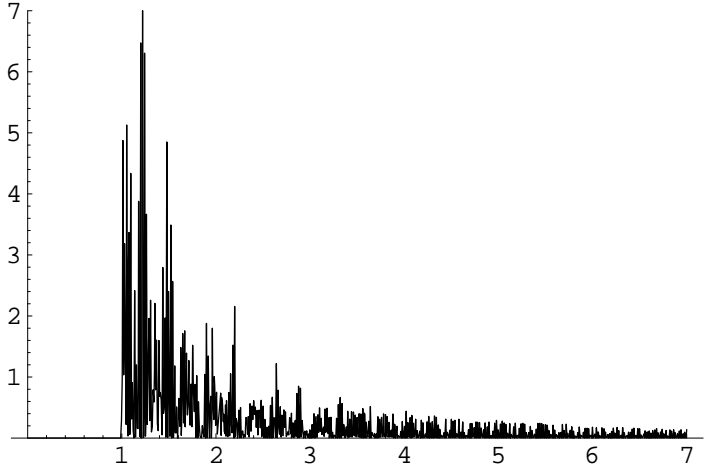}
\caption{ \small The probability distribution (\ref{pppp}) for
$\frac{\left||T_{k_1}| - |T_{k_2}| \right|}{\left||T_{k_1}| +
|T_{k_2}| \right|} = 0.7$ and $\sigma/q = 0.02$. The oscillations at
long time arise from the interference in the time-of-arrival due to
the different values of momentum.}
\end{figure}

To summarize: the condition $|T_{k_0}|^2 << 1$ is a prerequisite for
the existence of a meaningful decay law that does not depend on
detailed knowledge of the initial state. The condition $\sigma
\beta_{k_0}
>> 1$ can be relaxed substantially without loss of the exponential
decay phase. The regimes at which the condition $\sigma w_{k_0} <<
1$ is violated is expected to manifest the non-exponential decay
behavior most clearly.

\section{Comparison with the survival probability}

We finally compare our results with the estimations obtained from
the calculation of the survival probability for the same initial
states and potentials. For the initial state (\ref{initial}) we
obtain the following expression for the survival amplitude
\begin{eqnarray}
\langle \psi_0|e^{-i \hat{H}t} | \psi_0 \rangle =
\frac{1}{\sqrt{\pi} \sigma} \int_0^{\infty} dk \frac{|T_{k}|^2}{|1 +
R_{k}|^2} e^{-\frac{(k -k_0)^2}{\sigma^2} - i \frac{k^2}{2M}t}.
\label{ww}
\end{eqnarray}
In the above equation, we set with a good approximation the mode
normalization constant $|A_k|^2$ to $(2 \pi)^{-1}$ (the value for
the free particle on the half line). Expanding the logarithm of the
transmission and reflection amplitudes and keeping the lower order
terms, we obtain (still assuming that $\sigma /k_0 << 1$)
\begin{eqnarray}
\langle \psi_0|e^{-i \hat{H}t}|\psi_0 \rangle = \frac{1}{\sqrt{\pi}
\sigma} |T_{k_0}|^2 \sum_{n,m =0}^{\infty} (-\bar{R}_{k_0})^n
(-R_{k_0})^m
\nonumber \\
\times \int_{- \infty}^{\infty} dk e^{-\frac{(k -k_0)^2}{\sigma^2} -
i \frac{k^2}{2M}t} e^{i(m-n)\beta_{k_0} + 2 (w_{k_0} + \frac{n+m}{2}
s_{k_0})},
\end{eqnarray}
where $\beta_{k_0}, w_{k_0}, s_{k_0}$ are defined as previously. The
evaluation of the Gaussian integral yields
\begin{eqnarray}
\langle \psi_0|e^{-i \hat{H}t}|\psi_0 \rangle  = |T_{k_0}|^2 e^{-i
\frac{k_0^2}{2M}t} \sum_{n,m =0}^{\infty} (-\bar{R}_{k_0})^n
(-R_{k_0})^m
\nonumber \\
\times \exp \left\{ \frac{[i(\frac{k_0t}{M} - (m-n)\beta_{k_0}) + 2
(w_{k_0} + \frac{n+m}{2} s_{k_0})]^2}{4(\frac{1}{\sigma^2} +
i\frac{t}{2M})} \right\}. \label{oio}
\end{eqnarray}

In the regime that we obtained the exponential decay previously [$t
\sigma^2/M << 1, (w_{k_0} + \frac{n+m}{2} s_{k_0}) << 1$] this
expression yields
\begin{eqnarray}
\langle \psi_0|e^{-i \hat{H}t}|\psi_0 \rangle  = |T_{k_0}|^2 e^{-i
\frac{k_0^2}{2M}t} \sum_{n,m =0}^{\infty} (-\bar{R}_{k_0})^n
(-R_{k_0})^m e^{ - \frac{\sigma^2 \beta_{k_0}^2}{4} (m -n -
\frac{k_0 t}{M \beta_{k_0}})}.
\end{eqnarray}

If $\sigma \beta_{k_0} >> 1$, the Gaussian term has strong peaks at
times $t_{m, n}$ such that $ m  - n = \frac{k_0 t_{m,n}}{M
\beta_{k_0}}$. Ignoring the fine structure between such peaks, we
drop the summation over $m$ substituting $m = n +   \frac{k_0 t}{M
\beta_{k_0}}$. We then obtain
\begin{eqnarray}
\langle \psi_0|e^{-i \hat{H}t}|\psi_0 \rangle \simeq
 e^{-i \frac{k_0^2}{2M}t}
(-R_{k_0})^{k_0 t /M \beta_{k_0}}.
\end{eqnarray}

Hence the survival probability $w(t) = |\langle \psi_0|e^{-i
\hat{H}t}|\psi_0 \rangle|^2$ equals
\begin{eqnarray}
w(t) \simeq |R_{k_0}|^{\frac{2 k_0 t}{M \beta_{k_0}}}.
\end{eqnarray}
This expression describes exponential decay. Noting that our
derivation employed implicitly the condition $|T_{k_0}|^2 << 1$,
since we ignored the details of the distribution at  the microscopic
timescale $ M \beta_{k_0}/k_0$, we obtain
\begin{eqnarray}
w(t) \simeq e^{ - \frac{k_0 |T_{k_0}|^2}{M \beta_{k_0}} t}.
\label{wt}
\end{eqnarray}

We note that $w(t)$ decays exponentially with the same decay
constant $\Gamma$ that appears in Eq. (\ref{decay_rate}). Moreover,
the conditions for exponential decay are also the same.  This
agreement is quite remarkable given the very different characters of
the two objects: $w(t)$ is quadratic with respect to the initial
density matrix $\hat{\rho}$, while the probability $p(t)$ we
obtained is linear. We believe that this coincidence is a
consequence of the fact that the details of the initial state do not
affect the coarser behavior of the object constructed: after all,
$\Gamma$ only depends on the mean energy of the initial state and
not on any of its moments.

The agreement between the results from the study of the survival
probability and the detection probability only holds in the
exponential regime. For example, in the asymptotic regime of $t
\rightarrow \infty$, Eq. (\ref{ww}) yields
\begin{eqnarray}
\langle \psi_0|e^{- i \hat{H}t}|\psi_0 \rangle \sim t^{-(\frac{1}{2}
+ \alpha)},
\end{eqnarray}
where, as previously,  $\alpha$ is defined as $T_k \sim k^{\alpha}$
as $k \rightarrow 0$. Hence, $w(t) \sim t^{-(1 + 2\alpha)}$ and the
`decay probability' $-\dot{w}(t) \sim t^{-(2+2\alpha)}$, while as we
showed in Sec. 4.4, $p(t) \sim t^{-(3/2 + \alpha)}$.

More interesting is their divergence for the regimes that are not
characterized by an exponential decay phase. For the case of a decay
in which the condition $\sigma w_{k_0}$ is violated, we consider the
same initial state as in Eq. (\ref{superpos}). The assumption that
$\sigma/(k_1 - k_2) <<1$ leads to a suppression of the interference
terms in the evaluation of $\langle \psi_0|e^{-i
\hat{H}t}|\psi_0\rangle$. We finally obtain
\begin{eqnarray}
w(t) = \frac{1}{4} \left[ e^{- \Gamma_1t} + e^{- \Gamma_2t} + 2 e^{-
\sqrt{\Gamma_1 \Gamma_2}t}  \cos\left((\pi + \theta_{k_1})
\frac{k_1t}{M \beta_{k_1}} -  (\pi + \theta_{k_2} ) \frac{k_2t}{M
\beta_{k_2} } -  \frac{k_1^2 - k_2^2}{2M}t\right) \right], \;
\;\;\;\;\;
\end{eqnarray}
where $\theta_{k_{1,2}} = \mbox{Arg} R_{k_{1,2}}$.

 The expression above for $w(t)$  is different from Eq (\ref{ttt}) not only in the explicit
form and the persistence of  the oscillating term, but also in the
coefficient $\frac{1}{4}$ that appears in the diagonal terms. We
finally note that $w(t)$ is not a strictly decreasing function of
$t$, and for this reason $-\dot{w}(t)$ is non-positive and it cannot
be interpreted as a probability density for the time of decay. In
other words, outside the exponential regime one cannot trust the
results obtained from the survival amplitude to yield reliable
statistics for the measurement outcomes.

\section{Conclusions}

We reformulated tunneling as a time-of-arrival problem, in order to
study the decays of unstable states. We considered a general class
of potentials for a particle in the half-line. We saw that the
exponential regime is rather generic and we identified the specific
conditions that are necessary for its validity. This allowed us to
precisely identify the conditions necessary for the emergence of
decays that have no exponential phase.

The key feature of our construction is that there is neither
interpretational nor probabilistic ambiguity. The probabilities we
derive are obtained through a POVM, hence (unlike other approaches
to the problem) they are always positive and they respect the
convexity of the space of quantum states. The interpretation of
these objects is concretely operational, in the sense that it is
tied to the statistics for the measurement of particles' arrival
times.

Another important point is that the POVM we used in the derivation
of our results is defined in terms of the Hamiltonian, the initial
state and the location of the detector. It can therefore be applied
to much more general situations than the ones we considered here. In
particular, it can be used for the study of relativistic tunneling,
tunneling in open systems or even for the study of tunneling where
the barrier is not an external potential but is caused by
microscopic particle interaction (e.g. in nuclei).

Moreover, the analysis in Ref. \cite{AnSav06, AnSav07a} need not
apply only to particle systems. The measurement-theoretic context is
specified in a choice of the projectors that represent the type of
transition that is recorded by the measuring device \cite{AnSav06}.
These projectors can be completely general and they need not refer
to particle positions.  With a suitable choice for these objects,
the results can be generalized to field theoretic systems (e.g. for
the study of particle decays through field interactions) or even to
a cosmological setting.

Finally, we note that the derivation of the POVM and consequently of
these results depends crucially on concepts introduced by the
histories approach to quantum theory, in particular on the algebraic
(`logical') structure of the space of histories and on the
decoherence functional. We have found impossible to rephrase the
construction without explicitly referring the concepts above. For
this reason, we believe that these results provide an argument that
the distinctions and structures introduced by the histories approach
provide an extension to the quantum mechanical formalism with a
larger domain of applicability.

\section*{Acknowledgments}
I would like to thank N. Savvidou and D. Ghikas for useful
discussions on the issue.

\end{document}